\documentclass[pra,aps,amsmath,amssymb,showpacs]{revtex4}

\newcommand{\Tr}{\mathrm{Tr}}
\newcommand{\rh}{\mathrm{H}}
\newcommand{\id}{\mathrm{id}}
\newcommand{\hh}{\mathcal{H}}
\newcommand{\lnp}{\mathcal{L}}
\newcommand{\lsp}{\mathcal{L}_{+}}
\newcommand{\lpp}{\mathcal{L}_{++}}
\newcommand{\km}{\mathsf{K}}
\newcommand{\wm}{\mathsf{W}}
\newcommand{\am}{\mathsf{A}}
\newcommand{\ax}{\mathsf{X}}
\newcommand{\az}{\mathsf{Z}}
\newcommand{\as}{\mathsf{S}}
\newcommand{\xp}{\mathsf{P}}
\newcommand{\bsg}{\boldsymbol{\sigma}}
\newcommand{\bro}{\boldsymbol{\rho}}
\newcommand{\vro}{\boldsymbol{\varrho}}
\newcommand{\bor}{\boldsymbol{\omega}}
\newcommand{\U}{\widetilde{\mathsf{U}}}
\newcommand{\V}{\widetilde{\mathsf{V}}}
\newcommand{\D}{\mathsf{D}}
\newcommand{\pen}{\openone}
\newcommand{\eqs}{E_{q}^{(s)}}
\newcommand{\rqm}{\mathrm{E}_{q}^{(s)}}
\newcommand{\rmm}{\mathrm{M}_{q}^{(s)}}
\newcommand{\fec}{{\Phi}^{Q}}
\newcommand{\prq}{\psi^{QR}}
\newcommand{\rrg}{\boldsymbol{\rho}^{Q'R'}}
\newcommand{\hg}{\boldsymbol{\rho}^{Q}}
\newcommand{\rug}{\mathrm{G}}
\newcommand{\bts}{\widetilde{H}}

\begin{document}
\clearpage
\preprint{}

\title{Notes on entropic characteristics of quantum channels}
\author{Alexey E. Rastegin}
\affiliation{Department of Theoretical Physics, Irkutsk State University,
Gagarin Bv. 20, Irkutsk 664003, Russia}

\begin{abstract}
One of most important issues in quantum information theory
concerns transmission of information through noisy quantum
channels. We discuss few channel characteristics expressed by
means of generalized entropies. Such characteristics can often be
dealt in line with more usual treatment based on the von Neumann
entropies. For any channel, we show that the $q$-average output
entropy of degree $q\geq1$ is bounded from above by the
$q$-entropy of the input density matrix. Concavity properties of
the $(q,s)$-entropy exchange are considered. Fano type quantum
bounds on the $(q,s)$-entropy exchange are derived. We also give
upper bounds on the map $(q,s)$-entropies in terms of the output
entropy, corresponding to the completely mixed input.
\end{abstract}


\maketitle

\pagenumbering{arabic}
\setcounter{page}{1}

\section{Introduction}\label{sec1}

The quantum information theory treats quantum states and effects
as tools for information processing \cite{nielsen, bengtsson}.
Many of the important characteristics of quantum channels and
noise are expressed in terms of entropic measures. The Shannon
entropy of a classical probability distribution and the von
Neumann entropy of a density matrix are both of great importance.
In addition, some extensions of these entropic functionals were
found to be fruitful. Most important of them are the R\'{e}nyi
\cite{renyi} and Tsallis entropies \cite{tsallis}, the quantum
versions of which are frequently used as well.  A common treatment
of these and some other entropies in terms of unified
$(q,s)$-entropies was given in Ref. \cite{hey06}. For many values
of parameters $q$ and $s$, the quantum unified $(q,s)$-entropy
enjoys properties similarly to the standard von Neumann entropy
\cite{rastjst}. Entropic functionals have found a broad use in
studying features of quantum channels \cite{nielsen}.

In information-theoretical context, various applications of
quantum entropies are considered in Ref. \cite{bengtsson}. The
entropy exchange \cite{bsch96} and the map entropy \cite{zb04} are
widely known among different entropic characteristics. In Ref.
\cite{rprz12}, the uncertainty relations for a single quantum
operation are expressed in terms of the map and receiver
entropies. Entropic uncertainty relations are still the subject of
active research (see the reviews \cite{ww10, brud11} and
references therein). Additivity properties of map entropies with
respect to the tensor product of two channels are the subject of
active research \cite{rzf11}. For the minimum output entropy, this
question is regarded to be even more important
\cite{shor04,horod10}. Together with the von Neumann entropy, the
quantum R\'{e}nyi entropy has been applied for these aims. In Ref.
\cite{rast11a}, some properties of quantum channels have been
considered with use of unified entropies. In particular, the known
inequality of Lindblad \cite{glind91} with the entropy exchange
was extended to the case of unified entropies.

In this paper, we further study channel characteristics based on
generalized quantum entropies. The paper is organized as the
following: The preliminary material is reviewed in Section
\ref{sec2}. In Section \ref{sec3}, we derive an upper bound on the
$q$-average output entropy of degree $q\geq1$ in terms of the
corresponding entropy of the input. This bound, which holds for
arbitrary quantum channel, is explicitly shown with the
depolarizing channel. In Section \ref{sec4}, concavity properties
of the $(q,s)$-entropy exchange are examined. Fano type quantum
inequalities for the $(q,s)$-entropy exchange are obtained in
Section \ref{sec5}. These inequalities cover all the acceptable
values of the parameters $q$ and $s$. Upper bounds on the map
$(q,s)$-entropy in terms of the output entropy of completely mixed
input are presented in Section \ref{sec6}. In Section \ref{sec7},
we conclude the paper.

\section{Definition and notation}\label{sec2}

In this section, the preliminary material is given. Let
$\lnp(\hh)$ be the space of linear operators on $d$-dimensional
Hilbert space $\hh$. By $\lsp(\hh)$ and $\lpp(\hh)$, we
respectively denote the set of positive semidefinite operators
and the set of strictly positive ones. A density operator
$\bro\in\lsp(\hh)$ has unit trace, i.e., $\Tr(\bro)=1$. For
probability distribution $\{p_{j}\}$, the non-extensive entropy of
degree $q$ is defined as \cite{tsallis}
\begin{equation}
H_{q}(p_{j}):=\frac{1}{1-q}\left(\sum\nolimits_{j}p_{j}^{q}-1\right)=
-\sum\nolimits_{j}p_{j}^{q}{\,}\ln_{q}(p_{j})
\ . \label{tsend}
\end{equation}
Here the $q$-logarithm of argument $x>0$ is put for $q>0\neq1$ as
$\ln_{q}(x)=\bigl(x^{1-q}-1\bigr)/(1-q)$. In non-extensive
statistical mechanics, the entropy (\ref{tsend}) is commonly
referred to as Tsallis $q$-entropy. Up to a factor, the right-hand
side of Eq. (\ref{tsend}) was also put within information theoretic
context in Ref. \cite{HC67}. For any $p\in[0;1]$, the binary Tsallis
entropy is
\begin{equation}
\bts_{q}(p)=-p^{q}\ln_{q}(p)-(1-p)^{q}\ln_{q}(1-p)
\ . \label{btsdf}
\end{equation}
For physical quantity $A$, which takes the value $a_{j}$ with
probability $p_{j}$, the unnormalized $q$-average is defined as
\cite{curt91}
\begin{equation}
\langle{A}\rangle_{q}:=\sum\nolimits_{j}p_{j}^{q}{\,}a_{j}
\ . \label{unqa}
\end{equation}
Note that the formula (\ref{tsend}) can be rewritten as
$H_{q}(p_{j})=\bigl\langle-\ln_{q}(p_{j})\bigr\rangle_{q}$. The
R\'{e}nyi entropies form another important class of generalized
entropic functions. For $q>0\neq1$, the R\'{e}nyi $q$-entropy is
defined as \cite{renyi}
\begin{equation}
R_{q}(p_{j})=\frac{1}{1-q}{\>}\ln\left(\sum\nolimits_{j}p_{j}^{q}\right)
\ . \label{rend}
\end{equation}
When $q\to1$, the entropies (\ref{tsend}) and (\ref{rend}) both
recover the Shannon entropy
$H_{1}(p_{j})=-\sum_{j}p_{j}\ln{p}_{j}$. In quantum theory, the
above entropies have found use in studying measurement
uncertainties
\cite{rprz12,ww10,brud11,majer03,ww08,rast102,rast104,rast105,rastijtp}.
Applications of the entropy (\ref{tsend}) in analyzing
multiparticle production processes are discussed in Ref.
\cite{wilk11}.

For density matrix $\bro$, its von Neumann entropy is defined as
$\rh_{1}(\bro)=-\Tr(\bro\ln\bro)$. This is a quantum counterpart
of the Shannon entropy. Extensions of the von Neumann entropy are
obtained from Eqs. (\ref{tsend}) and (\ref{rend}) by replacing the
sums with the corresponding traces. The quantum $q$-entropy of
density operator $\bro$ is written as \cite{raggio}
\begin{equation}
\rh_{q}(\bro):=\frac{\Tr(\bro^q)-1}{1-q}
\ . \label{tsaeq}
\end{equation}
In Ref. \cite{hey06}, Hu and Ye proposed the class of more general
extensions of the standard entropic function. For positive
$q\neq1$ and any $s\neq0$, the unified $(q,s)$-entropy of
probability distribution $\{p_{j}\}$ is defined by
\begin{equation}
\eqs(p_{j}):=\frac{1}{(1-q){\,}s}{\,}\left[\left(\sum\nolimits_{j} p_{j}^{q}\right)^{{\!}s}-1\right]
\ . \label{undef}
\end{equation}
The quantum unified $(q,s)$-entropy of density operator $\bro$ is
defined as
\begin{equation}
\rqm(\bro):=\frac{1}{(1-q){\,}s}{\>}
\Bigl\{\bigl[\Tr(\bro^{q})\bigr]^s-1\Bigr\}
\label{qundef}
\end{equation}
for $q>0\neq1$ and any $s\neq0$ \cite{hey06}. In the limit
$s\to0$, the right-hand side of Eq. (\ref{qundef}) leads to the
quantum R\'{e}nyi entropy
\begin{equation}
{\mathrm{R}}_{q}(\bro):=\frac{1}{1-q}{\>}\ln\bigl[\Tr(\bro^{q})\bigr]
\ . \label{qredf}
\end{equation}
In the case $q=1$, we obtain the von Neumann entropy
$-\Tr(\bro\ln\bro)$. For the completely mixed state
$\bro_{*}=\pen/d$, the entropy (\ref{qundef}) reaches its maximum
\cite{hey06}
\begin{equation}
\rqm(\bro_{*})=\frac{1}{s}{\>}\ln_{q}\bigl(d^{s}\bigr)
\ . \label{qunma}
\end{equation}
For the Tsallis and Renyi entropies, we then have
$\rh_q(\bro_{*})=\ln_{q}(d)$ and
${\mathrm{R}}_{q}(\bro_{*})=\ln{d}$, respectively.

Consider a linear map $\Phi$ that takes elements of
${\mathcal{L}}(\hh)$ to elements of ${\mathcal{L}}(\hh')$, and
also satisfies the condition of complete positivity. Let $\id^{R}$
be the identity map on $\mathcal{L}(\hh_R)$, where the space
$\hh_R$ is assigned to auxiliary reference system. The complete
positivity implies that $\Phi\otimes\id^{R}$ transforms any
positive operator into a positive operator again for each
dimension of the extended space. Such linear maps are typically
called \emph{quantum operations} \cite{nielsen} or
\emph{super-operators} \cite{watrous1}. Each completely positive
map can be written in the operator-sum representation, namely
\begin{equation}
\Phi(\ax)=\sum\nolimits_{j}
\km_{j}{\,}\ax{\,}\km_{j}^{\dagger}
\ , \label{oper3}
\end{equation}
for all $\ax\in{\mathcal{L}}(\hh)$. Here the Kraus operators
$\km_{j}$ map the input space $\hh$ to the output space $\hh'$
\cite{nielsen,bengtsson}. If the map is trace-preserving, then the
Kraus operators satisfy the condition
\begin{equation}
\sum\nolimits_{j} \km_{j}^{\dagger}{\,}\km_{j}=\pen
\ , \label{tpcn}
\end{equation}
where $\pen$ is the identity in $\hh$. Trace-preserving quantum
operations are often called \emph{quantum channels} \cite{nielsen}.
Let $\bro$ be the input of quantum channel $\Phi$. The $j$th
effect occurs with the probability
\begin{equation}
p_{j}=\Tr\bigl(\km_{j}^{\dagger}{\,}\km_{j}{\,}\bro\bigr)
\ , \label{prj}
\end{equation}
and leads to the particular (normalized) output
\begin{equation}
\bro_{j}^{\prime}:=p_{j}^{-1}{\,}\km_{j}{\,}\bro{\,}\km_{j}^{\dagger}
\ . \label{rhpj}
\end{equation}
Since an informational content of quantum states is measured by
means of one or another entropies, various relations between input
and output entropies are of great importance. In the following,
inequalities of such a kind are formulated in terms of Tsallis
$q$-entropies and unified $(q,s)$-entropies.

The Jamio{\l}kowski--Choi representation \cite{jam72,choi75} is
another convenient description for completely positive maps. Let
$\hh_Q=\hh_R=\hh$ and $\{|\nu\rangle\}$ be an orthonormal basis in
$\hh$. To this basis we assign the normalized pure state
\begin{equation}
|\phi_{+}\rangle:=\frac{1}{\sqrt{d}}{\,}\sum_{\nu=1}^{d}
{|\nu\rangle\otimes|\nu\rangle} \ , \label{phip}
\end{equation}
where $d=\mathrm{dim}(\hh)$. One introduces the operator
$\bsg(\Phi):=\Phi\otimes\id\bigl(|\phi_{+}\rangle\langle\phi_{+}|\bigr)$,
acting on the space $\hh^{\otimes2}$. The matrix
$\D(\Phi)=d{\,}\bsg(\Phi)$ is usually called \emph{dynamical matrix} or
\emph{Choi matrix} \cite{zb04}. For each
$\ax\in{\mathcal{L}}(\hh)$, the action of the map $\Phi$
can be recovered from $\D(\Phi)$ by means of the relation
\cite{watrous1}
\begin{equation}
\Phi(\ax)=\Tr_{R}\left(\D(\Phi)(\pen\otimes\ax^{T})\right)
\ , \label{chjis}
\end{equation}
where $\ax^{T}$ denotes the transpose operator to $\ax$. The map
$\Phi$ is completely positive, if and only if the matrix
$\D(\Phi)$ is positive. The condition
$\Tr_{Q}\bigl(\D(\Phi)\bigr)=\pen$ is equivalent to that the map
$\Phi$ is trace-preserving and the rescaled matrix $\bsg(\Phi)$
is of unit trace. Writing positive semidefinite $\D(\Phi)$ as a
sum of one-rank matrices from $\lsp\left(\hh^{\otimes2}\right)$,
we can get a set of Kraus operators (for details, see Ref.
\cite{jam11} or section 5.2 in Ref. \cite{watrous1}). In the
following, the $(q,s)$-entropy of a quantum channel is defined
in terms of its rescaled dynamical matrix.

\section{An upper bound on the $q$-average output entropy}\label{sec3}

In this section, we will prove an upper bound on the $q$-average
of quantum entropies of particular outputs (\ref{rhpj}). In
accordance with Eq. (\ref{unqa}), we define the $q$-average
entropy as the sum
\begin{equation}
\bigl\langle\rh_{q}(\bro_{j}^{\prime})\bigr\rangle_{q}:=\sum\nolimits_{j} p_{j}^{q}{\>}\rh_{q}(\bro_{j}^{\prime})
\ . \label{qent}
\end{equation}
This quantity is bounded from above as follows.

\newtheorem{t21}{Proposition}\label{qaeq}
\begin{t21}
For $q\geq1$, the $q$-average output entropy of arbitrary quantum
channel is bounded from above as
\begin{equation}
\bigl\langle\rh_{q}(\bro_{j}^{\prime})\bigr\rangle_{q}
\leq\rh_{q}(\bro)
\ . \label{qaeq1}
\end{equation}
\end{t21}

{\bf Proof.}
We take imaginary systems $R$ and $S$, both identical
to the original system $Q$. Consider an isometry
$\V:{\>}\hh_{Q}\rightarrow\hh_{Q}^{\prime}\otimes\hh_{R}\otimes\hh_{S}$,
which is defined as
\begin{equation}
\V{\,}|\psi\rangle=\sum\nolimits_{j} \km_{j}{\,}|\psi\rangle\otimes|\nu_{j}\rangle\otimes|\nu_{j}\rangle
\ . \label{isomdf}
\end{equation}
Here $\{|\nu_{j}\rangle\}$ is chosen orthonormal basis in
$\hh_{Q}$. To given input $\bro$ of the channel, we assign the
density operator
\begin{equation}
\bor^{Q^{\prime}RS}:=\V{\,}\bro{\,}\V^{\dagger}
=\sum_{ij} \km_{i}{\,}\bro{\,}\km_{j}^{\dagger}
\otimes|\nu_{i}\rangle\langle\nu_{j}|\otimes|\nu_{i}\rangle\langle\nu_{j}|
\ . \label{omedf}
\end{equation}
Following Ref. \cite{rfz10}, we put the notation
$\am_{ij}:=\km_{i}{\,}\bro{\,}\km_{j}^{\dagger}$, so that
$p_{j}=\Tr(\am_{jj})$ and $\bro_{j}^{\prime}=p_{j}^{-1}\am_{jj}$.
The corresponding reduced densities are written as
\begin{align}
&\bor^{Q^{\prime}S}=\sum\nolimits_{j} \am_{jj}\otimes|\nu_{j}\rangle\langle\nu_{j}|
\ , \label{omac}\\
&\bor^{R}=\sum\nolimits_{j} \Tr(\am_{jj})|\nu_{j}\rangle\langle\nu_{j}|
\ . \label{omb}
\end{align}
With respect to the chosen basis, the matrices
$\bor^{Q^{\prime}S}$ and
$\bor^{R}={\mathrm{diag}}\bigl[p_{j}\bigr]$ have a diagonal form.
Let $\lambda_{k}^{(j)}$ denote eigenvalues of the positive
operator $\am_{jj}$; then
\begin{equation}
p_{j}=\sum\nolimits_{k} \lambda_{k}^{(j)}
\ , \label{plam}
\end{equation}
and the spectrum of $\bro_{j}^{\prime}$ contains
$p_{j}^{-1}\lambda_{k}^{(j)}$. The $q$-entropies of the densities
(\ref{omac}) and (\ref{omb}) are respectively expressed as
\begin{equation}
\rh_{q}\bigl(\bor^{Q^{\prime}S}\bigr)=H_{q}\bigl(\lambda_{k}^{(j)}\bigr)
\ , \qquad
\rh_{q}\bigl(\bor^{R}\bigr)=H_{q}(p_{j})
\ . \label{hqoom}
\end{equation}
Due to the definition (\ref{tsend}) and the relations
(\ref{hqoom}), the quantity (\ref{qent}) is then
rewritten as
\begin{align}
\sum\nolimits_{j} p_{j}^{q}
{\ }\frac{1}{1-q}\left[\sum\nolimits_{k}\biggl(\frac{\lambda_{k}^{(j)}}{p_{j}}\biggr)^{q}-1\right]
&=\frac{1}{1-q}{\>}\sum\nolimits_{jk}\bigl(\lambda_{k}^{(j)}\bigr)^{q}
-\frac{1}{1-q}{\>}\sum\nolimits_{j}p_{j}^{q}
\nonumber\\
&=\rh_{q}\bigl(\bor^{Q^{\prime}S}\bigr)-\rh_{q}\bigl(\bor^{R}\bigr)
\leq\rh_{q}\bigl(\bor^{Q^{\prime}RS}\bigr)
\ . \label{trinen}
\end{align}
The last step is the triangle inequality for the quantum
$q$-entropy of degree $q\geq1$ \cite{rastfn}. It directly follows
from the subadditivity property, which has been conjectured in Ref.
\cite{raggio} and later proved in Ref. \cite{auden07}. Since the
transformation $\V$ is an isometry, the three-partite state
(\ref{omedf}) and the input state $\bro$ have the same non-zero
eigenvalues \cite{rfz10}. Hence the entropies of these states
concur. $\blacksquare$

The $q$-average of $q$-entropies of particular outputs
$\bro_{j}^{\prime}$ is bounded from above by the input
$q$-entropy. The relation (\ref{qaeq1}) can be treated as an
extension of the second bound in theorem 1 of Ref. \cite{rfz10}.
That bound on the average output entropy is yielded from Eq.
(\ref{qaeq1}) by taking $q=1$. Incidentally, the first inequality
of theorem 1 in Ref. \cite{rfz10} states that the Holevo quantity
is not larger than the exchange entropy. The strong subadditivity
of the von Neumann entropy plays a principal role in its proof. In
classical regime, the $q$-entropy does enjoy the strong
subadditivity for $q>1$ \cite{sf06}, however this is an open
question for quantum $q$-entropies.

We now illustrate the above result with the depolarizing channel.
This well-known model represents a decohering qubit
\cite{preskill}. In the example, we will denote the identity
$2\times2$-matrix by $\pen$ and the usual Pauli matrices by
$\bsg_{x}$, $\bsg_{y}$, and $\bsg_{z}$, respectively. The Kraus
operators can be written as \cite{preskill}
\begin{equation}
\km_{0}=\sqrt{1-p}{\>}\pen
{\>}, \qquad
\km_{j}=\sqrt{\frac{p}{3}}{\>}\bsg_{j}
{\>}, \label{kedc}
\end{equation}
where $j=x,y,z$. Here, the parameter $p\in[0;1]$ characterizes a
probability that an error occurs. Since the case $p=0$ gives
the identity map, we will further take $p\neq0$. In terms of the
Bloch vector $\vec{s}=(s_{x},s_{y},s_{z})$, the input density
matrix is written as
\begin{equation}
\bro=\frac{1}{2}{\>}\bigl(\pen+\vec{s}\cdot\vec{\bsg}\bigr)
{\>}. \label{blor}
\end{equation}
Its eigenvalues are $\bigl(1\pm|\vec{s}|\bigr)/2$. For impure
states, we have $|\vec{s}|<1$. The particular outputs of the
channel are the following. First, one gives
$\bro_{0}^{\prime}=\bro$. For $j=x,y,z$, the Bloch vector of
$\bro_{j}^{\prime}$ has the same $j$-component as the input one,
whereas other two components are with reversed sign. The four
particular outputs have the same eigenvalues as the input $\bro$,
whence $\rh_{q}(\bro_{j}^{\prime})=\rh_{q}(\bro)$ for all $j$.
Further, the probabilities are $p_{0}=1-p$ and $p_{j}=p/3$ for
$j=x,y,z$. With impure $\bro$, the relation (\ref{qaeq1}) is then
reduced to
\begin{equation}
f_{q}(p):=(1-p)^{q}+3^{1-q}p^{q}\leq1
\ . \label{blor1}
\end{equation}
When $q\geq1$, the inequality (\ref{blor1}) actually holds for all
$p\in[0;1]$. Moreover, for $q>1$ and $p\in(0;1]$ the quantity
$f_{q}(p)$ is strictly less than $1$. That is, for all impure
inputs and non-zero $p$ we obtain
\begin{equation}
\bigl\langle\rh_{q}(\bro_{j}^{\prime})\bigr\rangle_{q}<\rh_{q}(\bro)
\ . \label{blor2}
\end{equation}
The last follows from the convexity of the function
$p\mapsto{f}_{q}(p)$, $f_{q}(0)=1$ and $f_{q}(1)=3^{1-q}<1$.
We have observed a simple test for the validity of Eq.
(\ref{qaeq1}). The depolarizing channel gives also an example, in
which the relation (\ref{qaeq1}) is obviously violated with
$0<q<1$. Here, the function $p\mapsto{f}_{q}(p)$ is concave,
$f_{q}(0)=1$ and $f_{q}(1)=3^{1-q}>1$, whence $f_{q}(p)>1$ for
each $p\in(0;1]$. Thus, for all impure inputs and non-zero $p$ we
should rewrite the inequality (\ref{blor2}) in opposite direction.
To this point, we recall that the triangle inequality for the
quantum $q$-entropy of degree $q\geq1$ has been crucial in the
proof of Eq. (\ref{qaeq1}).

\section{Concavity properties of the $(q,s)$-entropy exchange}\label{sec4}

In this section we study some properties of the $(q,s)$-entropy
exchange, which were introduced in Ref. \cite{rast11a}. To do so,
we explicitly separate the principal system $Q$ from the imaginary
reference system $R$ and environment $E$. The Hilbert spaces are
denoted by $\hh_{Q}$, $\hh_{R}$, and $\hh_{E}$, respectively.
Under the action of quantum channel $\fec$, the input state $\hg$
of system $Q$ is transformed into the output state $\fec(\hg)$. To
see the entanglement transmission, we consider a purification
$|\prq\rangle\in\hh_{Q}\otimes\hh_{R}$, which is mapped into the
final density matrix
\begin{equation}
\rrg=\fec\otimes\id^{R}\left(|\prq\rangle\langle\prq|\right)
\ . \label{strqe}
\end{equation}
The state of the system $R$ itself is not changed, i.e.,
$\Tr_{Q}\bigl(\rrg\bigr)=\Tr_{Q}\left(|\prq\rangle\langle\prq|\right)$.
The entanglement fidelity is defined as \cite{bsch96}
\begin{equation}
F\bigl(\hg,\fec\bigr):=\langle\prq|\rrg|\prq\rangle
\ . \label{efid}
\end{equation}
The notion of entropy exchange is very important, since it is
related to channel capacity. In our notation, the von Neumann
entropy exchange is put as
\begin{equation}
\rh_{1}\bigl(\rrg,\fec\bigr):=-\Tr\bigl(\rrg\ln\rrg\bigr)
\ . \label{vneex}
\end{equation}
This quantity was independently introduced by Lindblad
\cite{glind91} and Schumacher \cite{bsch96}. Putting an
environment $E$, we have an environmental representation of the
quantum channel $\fec$. It is written in terms of unitary operator
$\U$ on $\hh_{E}\otimes\hh_{Q}$ and pure state
$|e_{0}\rangle\in\hh_{E}$ as
\begin{equation}
\fec(\hg)=\Tr_{E}\left\{\U\bigr(|e_{0}\rangle\langle{e}_{0}|\otimes\hg\bigr)\U^{\dagger}\right\}
\ , \label{quopc}
\end{equation}
where the partial trace is taken over an environment. As the final
state
$\bigl(\U\otimes\pen^{R}\bigr)|e_{0}\rangle\otimes|\prq\rangle$ of
the system $EQR$ is pure, the output matrices
$\bro^{E'}\in\lsp(\hh_{E})$ and
$\rrg\in\lsp\bigl(\hh_{Q}\otimes\hh_{R}\bigr)$ have the same
non-zero eigenvalues. So the right-hand side of Eq. (\ref{vneex})
is equal to the von Neumann entropy of the output state
$\bro^{E'}$. Using the entropic functional (\ref{qundef}), we
define the $(q,s)$-entropy exchange as \cite{rast11a}
\begin{equation}
\rqm\bigl(\hg,\fec\bigr):=\rqm\bigl(\rrg\bigr)=\rqm\bigl(\bro^{E'}\bigr)
\ . \label{qseedf}
\end{equation}
This functional quantifies an amount of the $(q,s)$-entropy
introduced by the channel $\fec$ into an initially pure
environment $E$. The final state of $E$ can be expressed as
\cite{nielsen}
\begin{equation}
\bro^{E'}=\sum\nolimits_{ij} w_{ij}{\,}|e_{i}\rangle\langle{e}_{j}|
\ , \label{reaf}
\end{equation}
where $w_{ij}=\Tr\bigl(\km_{i}{\,}\hg\km_{j}^{\dagger}\bigr)$ are
entries of the matrix $\wm=[[w_{ij}]]$. Using this matrix, the
$(q,s)$-entropy exchange (\ref{qseedf}) can be rewritten as
\begin{equation}
\rqm\bigl(\hg,\fec\bigr)=\frac{1}{(1-q){\,}s}{\>}
\Bigl\{\bigl[\Tr(\wm^{q})\bigr]^s-1\Bigr\}
\ . \label{wseedf}
\end{equation}
In a similar manner \cite{bsch96}, the entanglement fidelity is
represented by
$F\bigl(\hg,\fec\bigr)=\sum\nolimits_{j}\left|\Tr(\hg\km_{j})\right|^{2}$.
Since the main definitions are already given, we now simplify the
notation to $\rqm(\bro,\Phi)$ and $F(\bro,\Phi)$. We will omit the
superscript $Q$ whenever density matrices and quantum operations
are related to the principal system $Q$ only. We shall now pose
convexity or concavity properties of the $(q,s)$-entropy exchange
with respect to its arguments. A short comment should be made
here. For any $\ax\in\lsp(\hh)$ and non-zero real number $q$, one
defines the functional
\begin{equation}
\rug_{q}(\ax):=\bigl[\Tr(\ax^{q})\bigr]^{1/q}
\ . \label{gurdf}
\end{equation}
We claim that this functional satisfies
\begin{align}
\rug_{q}(\ax+\az)&\geq\rug_{q}(\ax)+\rug_{q}(\az) &(0<q<1)
\ , \label{mnkl}\\
\rug_{q}(\ax+\az)&\leq\rug_{q}(\ax)+\rug_{q}(\az) &(1<q<\infty) \
. \label{mnkg}
\end{align}
For $q=1$, we have
$\rug_{q}(\ax+\az)=\rug_{q}(\ax)+\rug_{q}(\az)$. The relations
(\ref{mnkl}) and (\ref{mnkg}) can be considered as a tracial
version of the classical Minkowski inequality. Since operators
$\ax$ and $\az$ do not generally commute, the Minkowski inequality
for tuples of numbers is not sufficient for justifying Eqs.
(\ref{mnkl}) and (\ref{mnkg}). The formula (\ref{mnkg}) is
actually no more than the triangle inequality for the Schatten
$q$-norm (see, e.g., sect. 7.1 of Ref. \cite{carlen09}). Assuming
$q\in(0;1)$, the inequality (\ref{mnkl}) can be regarded as the
super-additivity inequality for the Schatten $q$-anti-norm.
Anti-norms of positive matrices are extensively treated in the
paper \cite{bh10}. In principle, the relation (\ref{mnkl}) can be
obtained as a special case of proposition 3.7 of Ref. \cite{bh10}.
In Appendix \ref{trimy}, we give another proof, which is closely
related to the classical Minkowski inequality for number tuples.
Using Eqs. (\ref{mnkl}) and (\ref{mnkg}), the authors of Ref.
\cite{hey06} have examined concavity properties of the quantum
$(q,s)$-entropy. Note that the Minkowski inequality was used in
Ref. \cite{hey06} with no comments on the non-commutativity. It is
for this reason that we prove Eq. (\ref{mnkl}) in Appendix
\ref{trimy}.

\newtheorem{t31}[t21]{Proposition}
\begin{t31}\label{prop2}
Let $\Phi$, $\Psi$ be quantum channels, and let
$\bro,\vro\in\lsp(\hh)$ be density matrices. In the parameter
range
\begin{equation}
\Bigl\{(q,s):{\>}0<q\leq1,{\>}-\infty<s\leq{q}^{-1}\Bigr\}{\,}{\bigcup}
{\,}\Bigl\{(q,s):{\>}1\leq{q},{\>}{q}^{-1}\leq{s}<+\infty\Bigr\}
\ , \label{pdcon}
\end{equation}
for each $\theta\in[0;1]$, the $(q,s)$-entropy exchange satisfies
\begin{align}
 & \rqm\bigl(\theta\bro+(1-\theta)\vro,\Phi\bigr)\geq\theta{\,}\rqm(\bro,\Phi)+(1-\theta){\,}\rqm(\vro,\Phi)
\ , \label{facn}\\
& \rqm\bigl(\bro,\theta\Phi+(1-\theta)\Psi\bigr)\geq\theta{\,}\rqm(\bro,\Phi)+(1-\theta){\,}\rqm(\bro,\Psi)
\ . \label{sacn}
\end{align}
\end{t31}

{\bf Proof.} Summarizing the results of the works
\cite{hey06,rastjst}, we state the following. In the parameter
domain (\ref{pdcon}), for each $\theta\in[0;1]$ there holds
\begin{equation}
\rqm\bigl(\theta\bro+(1-\theta)\vro\bigr)\geq\theta{\,}\rqm(\bro)+(1-\theta){\,}\rqm(\vro)
\ . \label{edcon}
\end{equation}
Using Eqs. (\ref{mnkl}) and (\ref{mnkg}), the writers of Ref.
\cite{hey06} have proved this inequality in the domain
(\ref{pdcon}) for any $s\neq0$. The concavity for $0<q\leq1$ and
$s=0$, i.e., for Renyi's $q$-entropies of order $q\in(0;1]$, was
shown in theorem 2 of Ref. \cite{rastjst}.

Consider the inputs $\bro$ and $\vro$. An action of the
channel $\Phi$ results in the matrices $\wm_{\bro}$ and
$\wm_{\vro}$ with elements
$\Tr\bigl(\km_{i}\bro\km_{j}^{\dagger}\bigr)$ and
$\Tr\bigl(\km_{i}\vro\km_{j}^{\dagger}\bigr)$, respectively.
For the input $\theta\bro+(1-\theta)\vro$, the corresponding
matrix is written as
\begin{equation}
\wm=\theta{\,}\wm_{\bro}+(1-\theta){\,}\wm_{\vro}
\ . \label{wmrvr}
\end{equation}
Combining this with Eq. (\ref{wseedf}) and the property (\ref{edcon}),
we obtain the first claim (\ref{facn}). To prove Eq. (\ref{sacn}), we
recall the meaning of the systems $Q$ and $R$ from the definitions
(\ref{strqe})--(\ref{vneex}). By $\rrg_{\Phi}$ and $\rrg_{\Psi}$,
we respectively denote density operators generated by channels
$\Phi$ and $\Psi$ in the sense of Eq. (\ref{strqe}). By linearity, one
further obtains
\begin{equation}
\bigl(\theta\Phi+(1-\theta)\Psi\bigr)\otimes\id^{R}\left(|\prq\rangle\langle\prq|\right)
=\theta{\,}\rrg_{\Phi}+(1-\theta){\,}\rrg_{\Psi}
\ . \label{rgphp}
\end{equation}
Combining this with Eq. (\ref{qseedf}) and the property
(\ref{edcon}) finally leads to the second claim (\ref{sacn}).
$\blacksquare$

Proposition \ref{prop2} states that the $(q,s)$-entropy exchange
is concave in its first entry as well as in its second entry. For
the case $s=1$, i.e., for the Tsallis entropy exchange, these facts
were noted in Ref. \cite{rastfn}. Fano type inequalities have also
been obtained in that paper. Bounds of such a kind can also be
derived for the $(q,s)$-entropy exchange.

\section{Fano type inequalities for the $(q,s)$-entropy exchange}\label{sec5}

In this section, we use the results of Ref. \cite{rastfn} to
obtain upper estimates of the Fano type for the $(q,s)$-entropy
exchange with arbitrary values of the parameters. The following
statement takes place.

\newtheorem{t32}[t21]{Proposition}
\begin{t32}\label{prop3}
For $q>0$ and all real $s$, the $(q,s)$-entropy exchange is
bounded from above as
\begin{equation}
\rqm(\bro,\Phi)\leq\frac{1}{1-q}{\>}\ln_{1-s}\Bigl\{
1+(1-q)\bts_{q}\bigl(F(\bro,\Phi)\bigr)+(1-q)\bigl[1-F(\bro,\Phi)\bigr]^{q}{\,}\ln_{q}(d^2-1)
\Bigr\} \ . \label{eqsfin}
\end{equation}
\end{t32}

{\bf Proof.}
It has been shown in Ref. \cite{rastfn} that for all
$q>0$, there holds
\begin{equation}
\rh_{q}(\bro,\Phi)\leq\bts_{q}\bigl(F(\bro,\Phi)\bigr)+\bigl[1-F(\bro,\Phi)\bigr]^{q}{\,}\ln_{q}(d^2-1)
\ . \label{rstfr}
\end{equation}
In the limit $q=1$, this bound reduces to the standard quantum
Fano inequality. So we further assume $q\neq1$. Suppose also that
$s\neq0$. According to the definitions (\ref{tsaeq}) and
(\ref{qundef}), we can write
\begin{equation}
\rqm(\bro,\Phi)=\frac{1}{(1-q){\,}s}{\>}\Bigl\{\bigl[1+(1-q){\,}\rh_{q}(\bro,\Phi)\bigr]^{s}-1\Bigr\}
\ . \label{etcn}
\end{equation}
The function
$x\mapsto(1-q)^{-1}s^{-1}\Bigl\{\bigl(1+(1-q){\,}x\bigr)^{s}-1\Bigr\}$
is increasing for those values of $x$ that obey $1+(1-q){\,}x>0$.
Indeed, its derivative is then positive. Combining this with Eqs.
(\ref{rstfr}) and (\ref{etcn}) gives the relation (\ref{rstfr})
with $s\neq0$, since we have a functional identity
\begin{equation}
\frac{y^{s}-1}{(1-q){\,}s}=\frac{1}{1-q}{\>}\ln_{1-s}(y)
\end{equation}
In line with the definitions (\ref{tsaeq}) and (\ref{qredf}), we
further write
\begin{equation}
{\mathrm{R}}_{q}(\bro,\Phi)=\frac{1}{1-q}{\>}\ln\Bigl\{1+(1-q){\,}\rh_{q}(\bro,\Phi)\Bigr\}
\ . \label{rtcn}
\end{equation}
We now observe that the function
$x\mapsto(1-q)^{-1}\ln\bigl(1+(1-q){\,}x\bigr)$ monotonically
increases with values of $x$ such that $1+(1-q){\,}x>0$. Combining
this with Eqs. (\ref{rstfr}) and (\ref{rtcn}) leads to the
relation (\ref{rstfr}) with $s=0$. $\blacksquare$

For many values of the parameters, the above upper bounds can
slightly be simplified. In the parameter range
\begin{equation}
\Bigl\{(q,s):{\>}0<q\leq1,{\>}-\infty<s\leq1\Bigr\}{\,}{\bigcup}
{\,}\Bigl\{(q,s):{\>}1\leq{q},{\>}1\leq{s}<+\infty\Bigr\}
\ , \label{pdfn}
\end{equation}
there holds Eq. (\ref{rstfr}) with $\rqm(\bro,\Phi)$ instead of
$\rh_{q}(\bro,\Phi)$, namely
\begin{equation}
\rqm(\bro,\Phi)\leq\bts_{q}\bigl(F(\bro,\Phi)\bigr)+\bigl[1-F(\bro,\Phi)\bigr]^{q}{\,}\ln_{q}(d^2-1)
\ . \label{erstfr}
\end{equation}
The case $s=1$ is merely the bound (\ref{rstfr}) itself. So we
will prove Eq. (\ref{erstfr}) for $s\neq1$ (and also $q\neq1$).
Assume that $s\neq0$ as well. For positive $x$, we can write the
inequality
\begin{equation}
\frac{\bigl(1+(1-q){\,}x\bigr)^{s}-1}{(1-q){\,}s}=\int_{0}^{x}\bigl(1+(1-q){\,}t\bigr)^{s-1}dt
\leq\int_{0}^{x}dt=x
\ , \label{intin}
\end{equation}
whenever $\bigl(1+(1-q){\,}t\bigr)^{s-1}\leq1$. The latter holds
in the following two cases: (i) $q<1$ and $s<1$; (ii) $1<q$ and
$1<s$. Except for the values $q=1$ and $s=0,1$, these cases fully
cover the parameter range (\ref{pdfn}). Combining this with Eqs.
(\ref{rstfr}) and (\ref{etcn}) completes the proof of
(\ref{erstfr}) for $s\neq0,1$. The case $s=0$, when R\'{e}nyi's
entropy is dealt with, follows from Eq. (\ref{rtcn}) and the inequality
\begin{equation}
\frac{1}{1-q}{\>}\ln\bigl(1+(1-q){\,}x\bigr)=\int_{0}^{x}\frac{dt}{1+(1-q){\,}t}
\leq\int_{0}^{x}dt=x
\ , \label{intin0}
\end{equation}
which holds for $q<1$. So we obtain Eq. (\ref{erstfr}) for $0<q<1$
and $s=0$.

The formulas (\ref{eqsfin}) and (\ref{erstfr}) provide upper
estimates of the Fano type on the $(q,s)$-entropy exchange. These
estimates are expressed in terms of the entanglement fidelity and
the corresponding binary $q$-entropy. The following treatment can
be given here. The first summand in the right-hand side of Eq.
(\ref{erstfr}), i.e., the corresponding binary $q$-entropy, is
small when the entanglement fidelity is close to either zero or
one. But only for the former the second summand in the right-hand
side of Eq. (\ref{erstfr}) becomes almost maximal. When the
$(q,s)$-entropy exchange is large, the entanglement fidelity
should be small enough. That is, the entanglement between $Q$ and
$R$ has not been well preserved. On the other hand, if the channel
$\Phi$ does preserve the entanglement appropriately, then its
$(q,s)$-entropy exchange is small. The $(q,s)$-entropy
exchange succeeds to the standard entropy exchange.

\section{Upper bounds on the map $(q,s)$-entropy}\label{sec6}

In this section, we will further examine the map $(q,s)$-entropies
introduced in Ref. \cite{rast11a}. For quantum channel $\Phi$ with
the rescaled dynamical matrix $\bsg(\Phi)$, we define the map
$(q,s)$-entropy by
\begin{equation}
\rmm(\Phi):=\rqm\bigl(\bsg(\Phi)\bigr)
\ . \label{mapqs}
\end{equation}
This is an extension of the standard map entropy introduced in
Ref. \cite{zb04} as the von Neumann entropy of $\bsg(\Phi)$.
Essential properties of the standard map entropy were examined in
Ref. \cite{rzf11}. The map entropy is used to characterize the
decoherent behavior of given quantum channel. If the input of the
channel is taken to be completely mixed, then the entropy exchange
coincides with the map entropy \cite{rfz10}. The same can be shown
for the $(q,s)$-entropy exchange and the map $(q,s)$-entropy. Few
additivity properties of the map entropy (\ref{mapqs}) have been
analyzed in Ref. \cite{rast11a}. These results were based on some
extension of the Lindblad inequality \cite{glind91}. Recall that
the entropy exchange is bounded from above by the sum of the input
and output von Neumann entropies of the principal quantum system.
Hence, estimates on the output entropy of completely mixed input
state can be obtained. Such a question has arisen within studies
of so-called ``additivity conjecture'' concerning a product
quantum channel (for details, see the paper \cite{rzf11} and
references therein). A relevant extension of the Lindblad
inequality to the $(q,s)$-entropy entropy exchange is written as
follows. For $q>1$ and $s\geq{q}^{-1}$, we have \cite{rast11a}
\begin{equation}
\bigl|\rqm\bigl(\Phi(\bro)\bigr)-\rqm(\bro)\bigr|
\leq\rqm(\bro,\Phi)\leq
\rqm\bigl(\Phi(\bro)\bigr)+\rqm(\bro)
\ , \label{extt0}
\end{equation}
including permutations of the three entropies. The proof of Eq.
(\ref{extt0}) is based on the subadditivity of the quantum
$(q,s)$-entropy for $q>1$ and $s\geq{q}^{-1}$ \cite{rastjst}. The
latter has been obtained by a relevant extension of the reasons
given for quantum Tsallis' entropy in Ref. \cite{auden07}. Taking
the completely mixed state $\bro_{*}=\pen/d$ as an input, the
formula (\ref{extt0}) leads to an upper bound on the map
$(q,s)$-entropy, namely
\begin{equation}
\rmm(\Phi)\leq
\rqm\bigl(\Phi(\bro_{*})\bigr)+\frac{1}{s}{\>}\ln_{q}\bigl(d^{s}\bigr)
\ . \label{extt1}
\end{equation}
This bound is expressed in terms of the output entropy and the
dimensionality. It turns out that a stronger bound can be derived.
Moreover, it is applicable for arbitrary values of the parameters.
We have the following statement.

\newtheorem{t41}[t21]{Proposition}
\begin{t41}\label{prop4}
Let $\Phi$ be a quantum channel, and let $\bro_{*}$ be the completely
mixed state on $d$-dimensional space $\hh$. For $q>0$ and all real
$s$, the map $(q,s)$-entropy is bounded from above as
\begin{align}
 &\rmm(\Phi)\leq{d}^{(1-q)s}{\,}\rqm\bigl(\Phi(\bro_{*})\bigr)+\frac{1}{s}{\>}\ln_{q}\bigl(d^{s}\bigr)
\qquad (s\neq0)
\ , \label{rmsn}\\
 &\mathrm{M}_{q}^{(0)}(\Phi)\leq\mathrm{R}_{q}\bigl(\Phi(\bro_{*})\bigr)+\ln{d}
\ . \label{rms0}
\end{align}
\end{t41}

{\bf Proof.} In Ref. \cite{rcun12}, we have derived inequalities
between the unified entropies of a composite quantum system and
one of its subsystems. Namely, for the final density matrix
(\ref{strqe}) and all $q>0\neq1$ there holds
\begin{align}
 &\rqm\bigl(\rrg\bigr)\leq{d}^{(1-q)s}{\>}\rqm\bigl(\fec(\hg)\bigr)
+\frac{1}{s}{\>}\ln_{q}\bigl(d^{s}\bigr) \qquad (s\neq0)
\ , \label{rmsnn}\\
    &\mathrm{R}_{q}\bigl(\rrg\bigr)\leq\mathrm{R}_{q}\bigl(\fec(\hg)\bigr)+\ln{d}
\ . \label{rms00}
\end{align}
Here, we use the notation with superscripts of the systems $Q$ and
$R$. The inequalities (\ref{rmsnn}) and (\ref{rms00}) follow from
relations, which describe a change of some symmetric norms and
anti-norms under the operation of partial trace \cite{rcun12}.
Taking the completely mixed state as the channel input, the
left-hand sides of Eqs. (\ref{rmsnn}) and (\ref{rms00}) are equal
to the corresponding map entropies. Omitting the superscripts, we
have arrived at the claims (\ref{rmsn}) and (\ref{rms0}).
$\blacksquare$

The inequalities (\ref{rmsn}) and (\ref{rms0}) give upper bounds
on the map $(q,s)$-entropies for all considered values of the
parameters $q$ and $s$. The formula (\ref{extt1}) provides an
upper bound only for $q>1$ and $s\geq{q}^{-1}$. In this case, the
power exponential $(1-q)s$ in the right-hand side of Eq.
(\ref{rmsn}) is negative. The new bound (\ref{rmsn}) is
stronger than Eq. (\ref{extt1}). In the standard limit $q\to1$,
these bounds obviously coincide. Similar bounds in terms of the
output $(q,s)$-entropies have been presented in Ref.
\cite{rcun12}. Replacing the dimensionality with the rank of the
dynamical matrix and the $\bro_{*}$ with arbitrary input $\bro$,
the right-hand sides of (\ref{rmsn}) and (\ref{rms0}) provide
upper bounds on the corresponding input entropies. All the bounds
follow from inequalities relating certain norms and anti-norms
before and after partial trace \cite{rcun12}. We may use the
results (\ref{rmsn}) and (\ref{rms0}) for estimating the map
$(q,s)$-entropies from above, when the corresponding output
entropies are known, exactly or approximately.

\section{Conclusion}\label{sec7}

In this paper, we have discussed some channel characteristics
expressed in terms of generalized quantum entropies. Generally,
discussed properties are dependent on the parameter values. In
many respects, generalized-entropy characteristics can be treated
similarly to the ones based on the von Neumann entropy. It turned
out that for any quantum channel and $q\geq1$, the $q$-average of
$q$-entropies of particular outputs is bounded from above by the
corresponding input entropy. This result is an extension of the
bound given for the von Neumann output and input entropies in Ref.
\cite{rfz10}. In a wide range of parametric values, the
$(q,s)$-entropy exchange enjoys concavity in any of its two
entries. Using Fano type quantum inequality of Ref. \cite{rastfn},
similar estimates on the $(q,s)$-entropy exchange have been
derived for all acceptable values of the parameters. We also obtain
some upper bounds on the map $(q,s)$-entropy introduced in Ref.
\cite{rast11a}. Thus, the presented results can be regarded as a
supplement and development of the previously given facts on
entropic characteristics of quantum channels.

\acknowledgments
The author is grateful to Karol \.{Z}yczkowski for helpful correspondence.

\appendix

\section{A tracial inequality of Minkowski type}\label{trimy}

In this section, we will prove Eq. (\ref{mnkl}) on base of the
classical Minkowski inequality. For strictly positive operators,
we can also allow negative exponential $q<0$. Let $x$ be a
$d$-tuple of strictly positive numbers. For $q\neq0$, we introduce
the function
\begin{equation}
G_{q}(x):=\Bigl(\sum\nolimits_{j=1}^{d}x_{j}^{q}\Bigr)^{1/q}
\ . \label{gqxdf}
\end{equation}
We can write $\rug_{q}(\ax)=G_{q}\bigl(\lambda(\ax)\bigr)$,
where $\lambda(\ax)$ denotes a $d$-tuple of positive eigenvalues
of $\ax$. For $q\geq1$, the right-hand side of Eq. (\ref{gqxdf})
gives the symmetric gauge function assigned to the Schatten
$q$-norm. The function (\ref{gqxdf}) clearly obeys permutation
symmetry and homogeneity, i.e., $G_{q}(\alpha{x})=\alpha{G}_{q}(x)$
with factor $\alpha>0$. In line with the usual Minkowski
inequality (see, e.g., theorem 25 in Ref. \cite{hardy}), for
$q<1\neq0$ there holds
\begin{equation}
G_{q}(x+y+\ldots+z)\geq{G}_{q}(x)+G_{q}(y)+\ldots+G_{q}(z)
\ . \label{gxydf}
\end{equation}
The cases of equality in Eq. (\ref{gxydf}) and more details are
provided in Ref. \cite{hardy}. The following statement takes place.

\newtheorem{aa1}[t21]{Proposition}
\begin{aa1}\label{proa1}
Let $\hh$ be finite-dimensional Hilbert space and
$\ax,\az\in\lpp(\hh)$. For all non-zero $q<1$, the functional
(\ref{gurdf}) satisfies
\begin{equation}
\rug_{q}(\ax+\az)\geq\rug_{q}(\ax)+\rug_{q}(\az)
\ . \label{ngkl}
\end{equation}
\end{aa1}

{\bf Proof.} Due to the Ky Fan maximum principle \cite{kyfan}, the
vector $\lambda(\ax+\az)$ is majorized by
$\lambda(\ax)+\lambda(\az)$. Namely, for all $k=1,\ldots,d$ there
holds
\begin{equation}
\sum\nolimits_{j=1}^{k}\lambda_{j}(\ax+\az)^{\downarrow}\leq
\sum\nolimits_{j=1}^{k}\lambda_{j}(\ax)^{\downarrow}+\sum\nolimits_{j=1}^{k}\lambda_{j}(\az)^{\downarrow}
\ , \label{kyfpr}
\end{equation}
with the equality $\Tr(\ax+\az)=\Tr(\ax)+\Tr(\az)$. Here the
arrows down imply that the eigenvalues should be put in the
decreasing order. It is well known that there exists a doubly
stochastic matrix $\as$ such that
\begin{equation}
\lambda(\ax+\az)^{\downarrow}=\as
\bigl[\lambda(\ax)^{\downarrow}+\lambda(\az)^{\downarrow}\bigr]
\ . \label{sdsm}
\end{equation}
For details, see point 4.3.33 of Ref. \cite{hjhon85}. The Birkhoff
theorem says that each doubly stochastic matrix is a convex
combination of finitely many permutation matrices \cite{hjhon85},
that is
\begin{equation}
\as=\sum\nolimits_{i}\alpha_{i}{\,}\xp_{i}
\ , \qquad \alpha_{i}\geq0
\ , \qquad \sum\nolimits_{i}\alpha_{i}=1
\ . \label{btcc}
\end{equation}
Combining this fact with the inequality (\ref{gxydf}) and using
both the homogeneity and symmetry, we then obtain
\begin{align}
&G_{q}\Bigl(\lambda(\ax+\az)^{\downarrow}\Bigr)=
G_{q}\Bigl(\sum\nolimits_{i}\alpha_{i}{\,}\xp_{i}\bigl[\lambda(\ax)^{\downarrow}+\lambda(\az)^{\downarrow}\bigr]\Bigr)
\nonumber\\
&\geq\sum\nolimits_{i}\alpha_{i}{\,}G_{q}\Bigl(\xp_{i}\lambda(\ax)^{\downarrow}\Bigr)
+\sum\nolimits_{i}\alpha_{i}{\,}G_{q}\Bigl(\xp_{i}\lambda(\az)^{\downarrow}\Bigr)
\nonumber\\
&=\sum\nolimits_{i}\alpha_{i}{\,}G_{q}\Bigl(\lambda(\ax)^{\downarrow}\Bigr)+
\sum\nolimits_{i}\alpha_{i}{\,}G_{q}\Bigl(\lambda(\az)^{\downarrow}\Bigr)
\ . \label{cmbpp}
\end{align}
The right-hand sides of Eqs. (\ref{ngkl}) and (\ref{cmbpp}) are equal
in view of $\sum_{i}\alpha_{i}=1$. $\blacksquare$

\end{document}